\documentclass[a4paper,11pt]{article}
\pdfoutput=1

\usepackage{jcappub} 

\makeatletter
\gdef\@fpheader{}
\makeatother

\newcommand{\bmt}{\begin{pmatrix}}
\newcommand{\emt}{\end{pmatrix}}
\newcommand{\ba}{\begin{array}{c}}
\newcommand{\ea}{\end{array}}
\newcommand{\be}{\begin{equation}}
\newcommand{\ee}{\end{equation}}
\newcommand{\bea}{\begin{eqnarray}}
\newcommand{\eea}{\end{eqnarray}}

\newcommand{\bi}{\begin{itemize}}
\newcommand{\ei}{\end{itemize}}

\newcommand{\baz}{\begin{array}{cc}}
\newcommand{\besub}{\begin{subequations}}
\newcommand{\eesub}{\end{subequations}}

\newcommand{\mathsym}[1]{{}}

\newcommand{\D}{\displaystyle}

\newcommand{\bt}{\begin{tabular}}
\newcommand{\et}{\end{tabular}}

\newcommand{\benu}{\begin{enumerate}}
\newcommand{\eenu}{\end{enumerate}}

\def\a{\alpha}

\def\g{\gamma}
\def\d{\delta}

\def\m{\mu}

\def\D{\Delta}
\def\L{\Lambda}

\def\q2 {q^2}

\def\bt{\begin{table}}
\def\et{\end{table}}

\def\ra{\rightarrow}

\usepackage[most]{tcolorbox}
\newtcbox{\mymath}[1][]{%
    nobeforeafter, math upper, tcbox raise base,
    enhanced, colframe=blue!30!black,
    colback=blue!30, boxrule=1pt,
    #1}

\usepackage{blindtext}

\begin{document}
\title{Relic neutrino decay solution to the excess radio background}
\author[a]{P.S. Bhupal Dev,}
\author[b,1]{Pasquale Di Bari, \note{Corresponding author.}}
\author[c]{Ivan Mart\'{i}nez-Soler}
\author[b]{and Rishav Roshan}

\affiliation[a]{Department of Physics and McDonnell Center for the Space Sciences, \\
Washington University, St. Louis, Missouri 63130, USA}
\affiliation[b]{School of Physics and Astronomy, University of Southampton,\\ Southampton, SO17 1BJ, U.K.}
\affiliation[c]{Institute for Particle Physics Phenomenology, Department of Physics, \\ Durham University, Durham DH1 3LE, U.K.}

\emailAdd{bdev@wustl.edu}
\emailAdd{p.di-bari@soton.ac.uk}
\emailAdd{ivan.j.martinez-soler@durham.ac.uk}
\emailAdd{r.roshan@soton.ac.uk}

\abstract{
The  excess radio background detected by ARCADE 2 represents a puzzle within the standard cosmological model. 
There is no clear viable astrophysical solution, and therefore, it might indicate the presence of new physics.  
Radiative decays of a relic neutrino $\nu_i$ (either $i=1$, or $i=2$, or $i=3$) into a sterile neutrino $\nu_{\rm s}$, assumed to be quasi-degenerate, provide  a solution that currently
evades all constraints posed by different cosmological observations and reproduces very well the ARCADE 2 data.  
We find  a very good fit to the ARCADE 2 data with best fit values $\tau_i = 1.46  \times 10^{21}\,{\rm s}$ and
$\Delta m_i = 4.0 \times 10^{-5}\,{\rm eV}$, where $\tau_i$ is the lifetime and $\Delta m_i$ is the mass difference between the 
decaying active neutrino and the sterile neutrino.  On the other hand, if relic neutrino decays do not explain ARCADE 2 data, then these
place a stringent constraint $\D m_i^{3/2} \tau_i \gtrsim 2 \times 10^{14}\,{\rm eV}^{3/2}\,{\rm s}$ in the range 
$1.4 \times 10^{-5} \, {\rm eV} < \D m_i < 2.5 \times 10^{-4}\,{\rm eV}$. 
The solution also predicts a stronger 
21 cm absorption global signal than the predicted one from the $\L$CDM model, with a contrast brightness temperature 
$T_{21} = -238^{+21}_{-20}\,{\rm mK}$ ($99\%$ C.L.) at redshift $z\simeq 17$. 
This is in mild tension with the even stronger signal found by the EDGES collaboration, $T_{21} = - 500^{+200}_{-500}\,{\rm mK} $, suggesting that this  might have been overestimated, possibly receiving a contribution from some unidentified foreground source.}

\maketitle

\section{Introduction}

The cosmological puzzles, in particular dark matter and matter-antimatter asymmetry of the universe, 
and the discovery of neutrino masses in neutrino oscillation experiments, clearly indicate the existence of new physics.  
However, we are still far from understanding the new framework that can accommodate them. Most of the efforts
have so far relied on collider physics, dark matter searches and low-energy neutrino experiments. However,
the rich variety of cosmological observations provides other interesting alternative ways to understand 
how new physics could manifest itself and these should not be overlooked or undermined. 

The Absolute Radiometer for Cosmology, Astrophysics and Diffuse Emission  (ARCADE 2) is a
balloon-borne instrument that has measured the absolute temperature of the sky in the 3--90 GHz radio frequency range \cite{Fixsen:2009xn}. 
A data analysis has found an extragalactic excess, in addition to the cosmic microwave background (CMB), at the low edge of the frequency range, approximately between 3 and 8 GHz. The excess fades away
at higher frequencies, so that results consistent with the CMB background spectrum measured by the Far Infrared Absolute Spectrophotometer (FIRAS) instrument  \cite{2002ApJFIRAS} are recovered in the far infrared frequency range $60$--$600$ GHz.  Moreover, when the ARCADE 2 data are combined with those from previous measurements at lower frequencies, in the range $22\,{\rm MHz}$--$1.42$ GHz,  it is found that a background power law spectrum with spectral index $\beta = -2.60 \pm 0.04$ 
describes well all data \cite{Fixsen:2009xn}. 

More recently, a study of the low frequency all-sky maps from the Long Wavelength Array (LWA)  measured the 
diffuse radio background between 40 and 80 MHz, also finding an excess \cite{Dowell:2018mdb}. 
In combination with the ARCADE 2 data,  this analysis confirms a background power law spectrum with spectral index $\beta = -2.58 \pm 0.05$ . This excess radio background cannot be explained by known population of sources since they give a contribution to the effective temperature that is 3--10 times smaller than the measured one \cite{Singal:2017jlh}.  
Part of the excess radio background is due to galactic synchrotron radiation but this galactic contamination 
is significantly below the background  \cite{Fornengo:2014mna}. 
Recently, low-redshift populations of discrete extragalactic radio sources have been also excluded by angular cross-correlating data of the diffuse radio sky with
matter tracers at different redshifts provided by galaxy catalogs and CMB lensing \cite{Todarello:2023nqd}. 
More exotic astrophysical explanations of the excess radio background  have proposed an origin from unknown radio source counts  at high redshift, such as supermassive black holes and star forming galaxies.  
However, a study of the diffuse extragalactic radio emission at 1.75 GHz using the Australia Telescope Compact  Array (ATCA), constrains the contribution to the background of  extended sources with angular size as large as 2 arcmin \cite{Vernstrom:2014uda}.  These and other observations place a strong upper limit on the  anisotropy of the excess radio background that is, therefore, extremely smooth \cite{Holder:2012nm}. This represents a strong constraint for an astrophysical origin. For this reason, many explanations in terms of new physics have been
proposed such as: WIMP annihilations or decays in extra-galactic halos \cite{Fornengo:2011cn}; 
synchrotron radiation emitted by the electrons produced by the late decays of a metastable particle in a magnetic field \cite{Cline:2012hb};
dark photons, produced by dark matter decays, oscillating into ordinary photons \cite{Caputo:2022keo};  superconducting cosmic strings \cite{Brandenberger:2019lfm,Cyr:2023yvj}; 
soft photon emission from accreting primordial black holes \cite{Mittal:2021dpe};
radiative decay of relic neutrinos into sterile neutrinos \cite{Chianese:2018luo}.
Also in this case the smoothness of the background strongly constrains some of these solutions, 
in particular those relating the origin of the excess radio background to dark matter.\footnote{We refer to dark matter as the dominant component of dark matter, cold or warm.}
 Some of these explanations are also constrained by other cosmological observations \cite{Acharya:2022txp,Acharya:2022vck}  but additional data will be necessary to draw firm conclusions. 
 In this respect, it is interesting that more measurements will come both from CMB experiments
 with a lower low frequency threshold than FIRAS, such as the Primordial Inflation Explorer (PIXIE) \cite{Kogut:2011xw} or
 its more advanced versions Super-PIXIE \cite{Kogut:2019vqh} and VOYAGE 2050 \cite{Chluba:2019nxa}, 
 that will measure deviations of the CMB spectrum from thermal down to 30 GHz,  and from radio interferometers  such as the Tenerife Microwave Spectrometer (TMS) that will cover the 10--20 GHz frequency range 
 \cite{2020SPIE11453E..0TR,Alonso-Arias:2021quq}.

An important point is that even if just a small fraction of the excess radio background detected at the present time has to be ascribed to redshifted radiation produced at high redshifts (at least $z \gtrsim 20$), this would produce a sizeable deviation of the cosmological 21 cm global absorption signal from the standard prediction \cite{Feng:2018rje},  where the only contribution comes from the CMB radiation  \cite{Furlanetto:2006jb}.  For example, even if only $0.1 \%$ of the  background has a high redshift origin, this would be enough to produce a detectable effect \cite{Feng:2018rje}.   Recently, it has been pointed out that a soft photon heating effect, due to inverse  bremsstrahlung absorption by the background electrons, should also be taken into account \cite{Acharya:2023ygd}. This can strongly reduce the absorption 21 cm global signal expected from the observed excess radio background and it has been specifically shown to be important in decaying particle scenarios, as the one we will consider. In any case some deviation from the standard prediction is expected.

It is then intriguing that the Experiment to Detect the Reionization Step (EDGES) collaboration has indeed claimed evidence of a 21 cm global absorption signal falling  approximately in the expected interval of redshifts within standard cosmology but about twice stronger \cite{Bowman:2018yin}. It has to be said that such a claim has not yet been confirmed by other experiments. Moreover, concerns related to various possible sources of underestimated systematic uncertainties have been discussed such as ionospheric effects \cite{Hills:2018vyr} and inhomogeneities of the ground below the EDGES antenna \cite{Bradley:2018eev}. In addition, the Shaped Antenna measurement of the Background RAdio Spectrum (SARAS 3) experiment collaboration has found no evidence of the absorption signal, rebutting the EDGES claim \cite{Singh:2021mxo}. Both concerns and the rebuttal have been addressed respectively 
in \cite{reply,Murray:2022uhg,Bevins:2022ajf}.  It is anyway important that a few experiments, in addition to SARAS 3, are underway to test the EDGES claimed anomaly in next years, including:
the Large-aperture Experiment to Detect the Dark Ages (LEDA) \cite{Bernardi:2016pva},
the Mapper of the IGM Spin Temperature (MIST) \cite{Singal:2022jaf},
Probing Radio Intensity at High-Z from Marion (PRIZM) \cite{PRIZM},
the Radio Experiment for the Analysis of Cosmic Hydrogen (REACH) \cite{deLeraAcedo:2022kiu}, and Hydrogen Epoch of Reionization Array (HERA)~\cite{HERA:2022wmy}.
Moreover, space and lunar-based global 21 cm experiments, such as 
Discovering Sky at the Longest wavelength (DSL) \cite{Chen:2020lok},  
the Lunar Surface Electromagnetic Experiment (LuSEE) \cite{2023arXiv230110345B}
and Probing ReionizATion of the Universe using Signal from Hydrogen (PRATUSH)~\cite{2023ExA}, 
are also planned to start operating in the next few years. These will be able to circumvent
observational systematics associated with the ionosphere and human-made radio frequency interference.

The anomalous absorption 21 cm signal claimed by EDGES has stimulated a variety of different 
proposals for its explanation. The signal can be expressed in terms of the 21 cm brightness temperature
$T_{21}(z)$. At redshifts $z \sim 10$--$25$ this is approximately proportional to  
$1 - T_{\gamma}(\nu_{21}^{\rm rest},z)/T_{\rm gas}(\nu_{21}^{\rm rest},z)$,  
where $T_{\gamma}(\nu_{21}^{\rm rest},z)$ is the effective temperature of
radiation and $T_{\rm gas}(\nu_{21}^{\rm rest},z)$ is the temperature of the gas at redshift $z$ 
at the 21 cm rest frequency $\nu_{21}^{\rm rest} = 1420 {\rm MHz}$. This corresponds today
to a redshifted (observed) frequency $\nu_{21}(z) = \nu_{21}^{\rm rest}/(1+z)$. 
Since in this range of redshifts one has $T_{\rm gas}(z) < T_{\gamma}(z)$, then $T_{21}(z) < 0$,
corresponding to an absorption signal.   The EDGES collaboration has claimed evidence of an absorption signal
just in the expected range of redshifts and with a minimum at $z_E \simeq 17.2$, corresponding to $\nu_{21}(z_E) \simeq 78\,{\rm MHz}$ but with a minimum value of $T_{21}$ about 2.5 times  lower than expected in  $\L$CDM.  Explanations of this anomalous discrepancy 
can be divided into two categories: (i) the  absorption signal is enhanced by cooling the gas temperature compared to the standard case; (ii) the radiation effective temperature  is higher than in the standard case, where it is given simply by the CMB temperature, implying the 
presence of some extra radiation in addition to relic thermal radiation that could also explain the excess radio background.

In the first category solutions necessarily rely on new physics. For example, it was proposed that gas is made colder by the interactions with the dark matter \cite{Barkana:2018lgd}. Different constraints single out a solution where the gas should interact with minicharged particles that give only a subpercent fraction of the total dark matter energy density \cite{Munoz:2018pzp}. In the second category one can think of astrophysical solutions relying on the existence of unknown  astrophysical sources of radiation in the radio frequencies.
For example, it has been proposed that accretion onto the first intermediate-mass black holes
between $z\simeq 30$ and $z\simeq 16$ can explain the EDGES anomaly \cite{Ewall-Wice:2018bzf}.
However, these kind of solutions necessarily rely on not so well established mechanisms that prevent X-ray emission with consequent  heating of the intergalactic medium that could change the reionization history incompatibly with the CMB anisotropy observations.  Because of the challenges encountered by astrophysical solutions able to explain the EDGES anomaly with 
the presence of extra radiation, also many solutions relying on new physics have been proposed, such as:
decay or de-excitations of some fraction of dark matter \cite{Fraser:2018acy};
resonant oscillations of dark photons into regular photons \cite{Pospelov:2018kdh};
radiative decay of relic neutrinos into sterile neutrinos \cite{Chianese:2018luo};
superconducting cosmic strings \cite{Brandenberger:2019lfm,Cyr:2023yvj};
conversion of mirror photons into standard photons \cite{AristizabalSierra:2018emu}.
In most cases
these have also been discussed in combination with an explanation of the ARCADE 2 excess radio background.
 In this paper we revisit the case of relic neutrino radiative  decays since it currently represents an attractive option 
 to produce extra non-thermal radiation for a few reasons: 
 \begin{itemize}
\item[(i)] It provides a solution to the ARCADE 2 excess and potentially also to the EDGES anomaly.
\item[(ii)]  It is minimal since it involves essentially only two parameters: the lifetime of neutrinos and the mass difference between the relic neutrinos and the invisible particle into which it decays. We will consider the case of a sterile neutrino as originally proposed in \cite{Chianese:2018luo} but it should be clear that the discussion might be extended to other cases. 
 \item[(iii)] The relic neutrino abundance can be assumed to be the standard one predicted by the $\L$CDM model
 and it is therefore fixed.
 \item[(iv)] Since neutrinos are fermions, their clustering is very limited, especially for current neutrino mass bound
 $m_{i} \lesssim 0.1\,{\rm eV}$, compared to cold dark matter \cite{Mertsch:2019qjv,Zimmer:2023jbb}.  Therefore,
 it predicts a very smooth excess radio background, as supported by the observations, and so it 
 does not suffer the tension of models where the excess radio background is related 
 to the dark matter abundance. 
 \end{itemize}

We will show that the spectrum predicted by this solution fits very well the ARCADE 2 data and has some very definite features
that makes it distinguishable from other models, like for example the single power law model proposed to explain also LWA data in a combined way. 
In particular, the spectrum presents a transition between two different power laws and has a sharp end point. It makes a very precise prediction
for the brightness contrast temperature $T_{21}$ in the 21 cm absorption global signal.  As we will see this is in mild tension with the EDGES anomaly
since it predicts a weaker signal, though still with a sizeable observable deviation from the standard case. We will also comment on the
fact that the same kind of spectrum can be obtained from dark matter decays/de-excitations but only for masses of dark matter below 10 keV.  

The paper is structured as follows. In Section~\ref{sec:2} we briefly review the derivation of the effective temperature of non-thermal radiation produced from radiative relic neutrino decays.  In Section~\ref{sec:3} we specialise it to zero redshift to fit the excess radio background data from ARCADE 2. We find the best fit and the $99 \%$ C.L. allowed region in the plane $\Delta m_1$ vs. $\tau_1$. In Section~\ref{sec:4} we consider the 21 cm cosmological absorption signal claimed by EDGES and also in this case we specialise the general formula and find the $99 \%$ C.L. allowed region noticing that there is a tension with the ARCADE 2 allowed region. We also derive the prediction on the brightness contrast temperature $T_{21}$ from ARCADE 2 data. 
In Section \ref{sec:5} we discuss the case of dark matter de-excitations, introducing a third parameter given by the dark matter mass and showing how, for dark matter masses much higher than about 10 keV, only the EDGES anomaly can be explained, not the ARCADE 2 excess radio background data. 
Finally, in Section~\ref{sec:6} we draw our conclusions making some final remarks.

\section{Effective temperature of non-thermal radiation}\label{sec:2}

Let us derive the effective  temperature of non-thermal photons produced
from the radiative decays of relic neutrinos. For definiteness,
we will refer to the decay of the lightest neutrino $\nu_1$, since this presents different attractive
features compared to the  decays of the two heavier neutrinos. However, our results can also
be extended to that case.  We will consider radiative two body decays into sterile neutrinos
with mass $m_{\rm s}$. The lightest active neutrino and the sterile neutrino are assumed to be quasi-degenerate,
so that $\Delta m_1 \equiv m_1 - m_{\rm s} \ll m_1$. We also assume that the lightest neutrinos decay non-relativistically 
in a way that at the production photons from decays are approximately monochromatic. 
In this case  the {\em specific intensity} at redshift $z$ of the non-thermal  photons of energy $E \leq \D m_1$
is given by\footnote{This expression further generalises the expression given in \cite{Chianese:2018luo} that extended the expression given in \cite{Masso:1999wj} for the case $z=0$.} 
\be\label{Inth}
I_{\gamma_{\rm nth}}(E,z)  = {1 \over 4 \pi} \, {d\varepsilon_{\g_{\rm nth}} \over d E}
\, = {n_{\nu_1}^{\infty}(z) \over 4\,\pi} 
{e^{-{t(a_{\rm D}) \over \tau_1}} \over H(a_{\rm D}) \, \tau_1} \,  ,
\ee
where $\tau_1$ is the active neutrino lifetime, $\varepsilon_{\g_{\rm nth}}$ is the energy density of the non-thermal radiation,
$H(a_{\rm D})$ and $t(a_{\rm D})$ are, respectively, the expansion rate and the age of the universe
calculated at the time of decay of the relic neutrinos that produced photons with 
energy $E$ at redshift $z$, corresponding to a redshift $z_{\rm D} \equiv a_{\rm D}^{-1} -1$  
and scale factor $a_{\rm D} = (E  /\D m_1)\, a \leq  a$, and
\be\label{nnu}
n^{\infty}_{\nu_1}(z) = {6\over 11}{\zeta(3)\over \pi^2}\,T^3(z)
\ee
is the relic neutrino number density at $z$ in the standard stable neutrino case ($T$ is the standard relic photon temperature).

The expansion rate at the decay, $H(a_{\rm D})$, can be calculated in the $\L$CDM model  as
\be
H(a_{\rm D}) = H_0\,\sqrt{\Omega_{{\rm M}0} \,a_{\rm D}^{-3} + \Omega_{\Lambda  0}} = 
H_0 \, \sqrt{\Omega_{M0}} \, a_{\rm D}^{-{3\over 2}} \, 
\left(1 + {a_{\rm D}^3 \over a_{\rm eq}^3} \right)^{{1 \over 2}}  \,  ,
\ee
where $a_{\rm eq} \equiv (\Omega_{M0}/\Omega_{\Lambda 0})^{1/3} \simeq 0.77$, $\Omega_{M0} \simeq 0.3111$, $H_0 \simeq t_0^{-1}$ and $t_0 \simeq 13.8 \, {\rm Gyr} \simeq 4.35 \times 10^{17}\,{\rm s}$ \cite{Planck:2018vyg}.\footnote{We can neglect statistical errors on cosmological parameters for the determination of 
the best fit value for $\tau_1$ (as a function of $\D m_1$). We just notice that 
taking $H_0 = t_0^{-1} = 71\,{\rm km}\,{\rm s}^{-1}\,{\rm Mpc}^{-1}$ corresponds to a kind of average value between the CMB determination 
$H_0 \simeq  67\,{\rm km}\,{\rm s}^{-1}\,{\rm Mpc}^{-1}$ \cite{Planck:2018vyg} and the SH0ES determination from SNIa 
$H_0 \simeq  73\,{\rm km}\,{\rm s}^{-1}\,{\rm Mpc}^{-1}$ \cite{Murakami:2023xuy}. 
This uncertainty on $H_0$, the so-called Hubble tension, 
is also negligible for the  determination of $\tau_1$ with current measurements of the excess radio background.} 
An analytical expression for the age of the universe at the decay, $t(a_{\rm D})$, 
can also be found within the $\L$CDM model \cite{DiBari:2018vba}:
\be
t(a_{\rm D}) = {2\over 3}\,{H_0^{-1}\over \sqrt{\Omega_{\Lambda 0}}}\,
\ln\left[\sqrt{\left({a_{\rm D}\over a_{\rm eq}}\right)^3} + \sqrt{1+\left({a_{\rm D}\over a_{\rm eq}}\right)^3} \right] \,  .
\ee

The specific intensity from non-thermal photons produced by 
relic neutrino decays would give,  at the frequency $\nu$ at redshift $z$, 
an excess with respect  to the specific intensity of the thermal radiation that is simply given by
\be
I_{\rm th}(E,z) \equiv {d{\cal F}^{\g_{\rm th}}_E \over dA\,dt \, dE \,d\Omega}  = {1 \over 4\pi} \, {d\varepsilon_{\gamma_{\rm th}} \over d E}
= {E^3 \over 4\,\pi^3}\,[e^{E/T(z)}-1]^{-1} \,  .
\ee
In the Rayleigh-Jeans tail, for $E = h \nu \ll T(z)$, the specific intensity depends linearly on temperature:
\be
I_{\rm th}(E,z) = {E^3 \over 4\,\pi^3}\,[e^{E/T(z)}-1]^{-1} \, \stackrel{E \ll T(z)}{\longrightarrow} \, {1 \over 4 \pi^3}\,T(z) \, E^2 \, . 
\ee
For some generic non-thermal radiation, it is customary to introduce an {\em effective photon temperature} $T_{\g_{\rm nth}}(E,z)$  defined as the temperature corresponding to a  thermal radiation with the same specific intensity at the given arbitrary frequency $\nu$ and redshift $z$, explicitly:
\be\label{Tgamma}
T_{{\gamma}_{\rm nth}}(E, z)=  {E \, \ln^{-1} \left(1+{E^3 \over 4\,\pi^3 \, I_{\g_{\rm nth}}(E,z)} \right)}   \,  .
\ee  
Again, in the low frequency limit, for $E \ll T_\g$, one has a 
simple linear relation between the effective temperature and the specific intensity:
\be\label{Tgammabis}
T_{{\gamma}_{\rm nth}}(E, z) \simeq   {4 \pi^3 \over E^2} \, I_{\g_{\rm nth}}(E,z) \,  .
\ee  
This gives us a general formula valid at any redshift. In the case of ARCADE 2 the measurements are made at zero redshift. In the
case of EDGES the `detection' occurs at $z \sim z_E$ and it is seen today as a (redshifted) absorption feature.

\section{Fitting the ARCADE 2 excess radio background}\label{sec:3}

Let us see how the expression (\ref{Inth}) can describe the ARCADE 2 measurements of the excess radio background 
temperature, defined as $T_{\rm ERB} \equiv T_{\gamma 0} -T_0$, shown in Table~\ref{table1}. 
Notice that the ARCADE 2 experiment found $T_0 = (2.729\pm 0.004)\,{\rm K}$, in very good
agreement with the FIRAS measurement.

\begin{figure}
\begin{center}
    \psfig{file=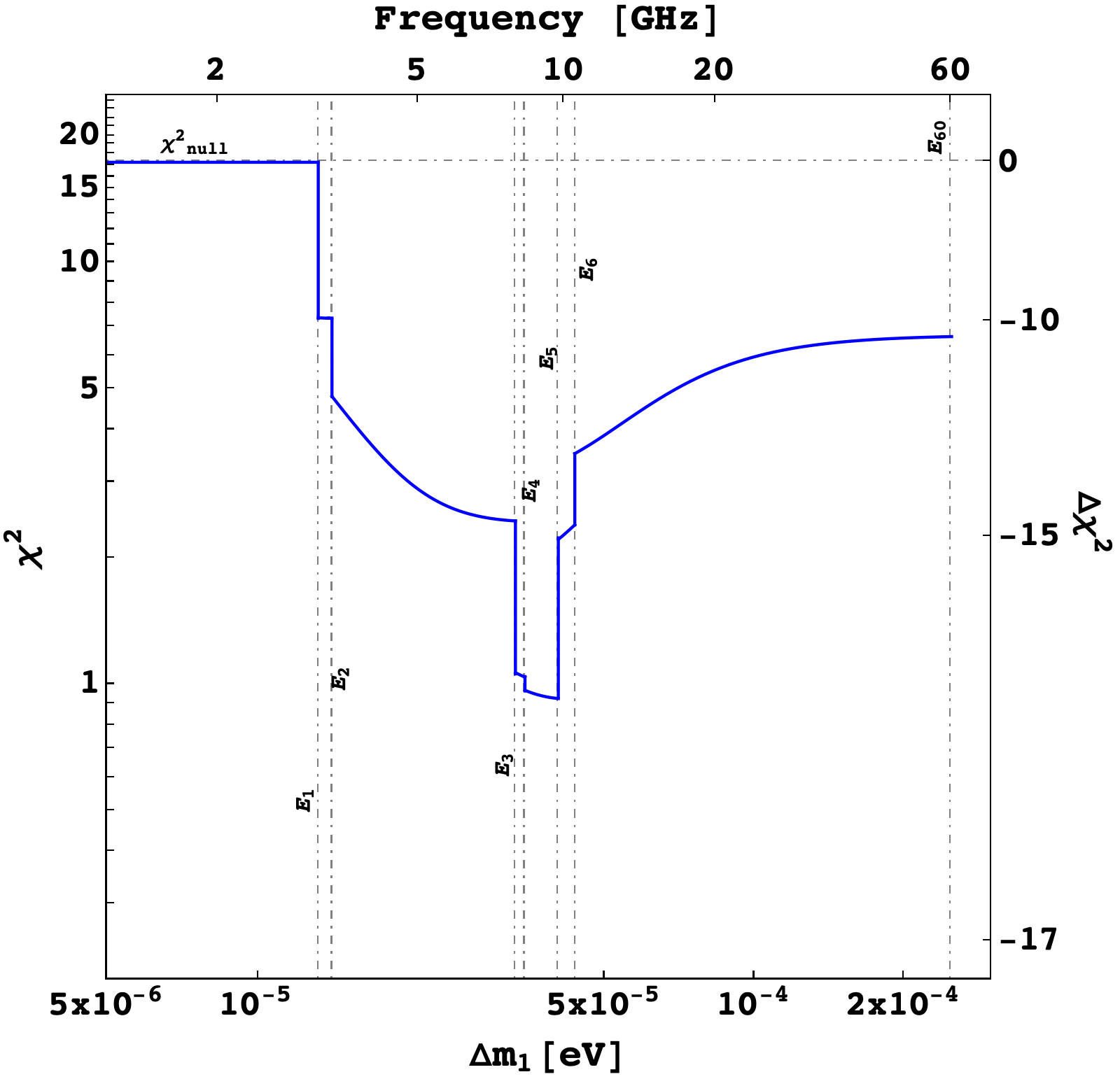,height=7.0cm,width=7.0cm} \hspace{4mm}
    \psfig{file=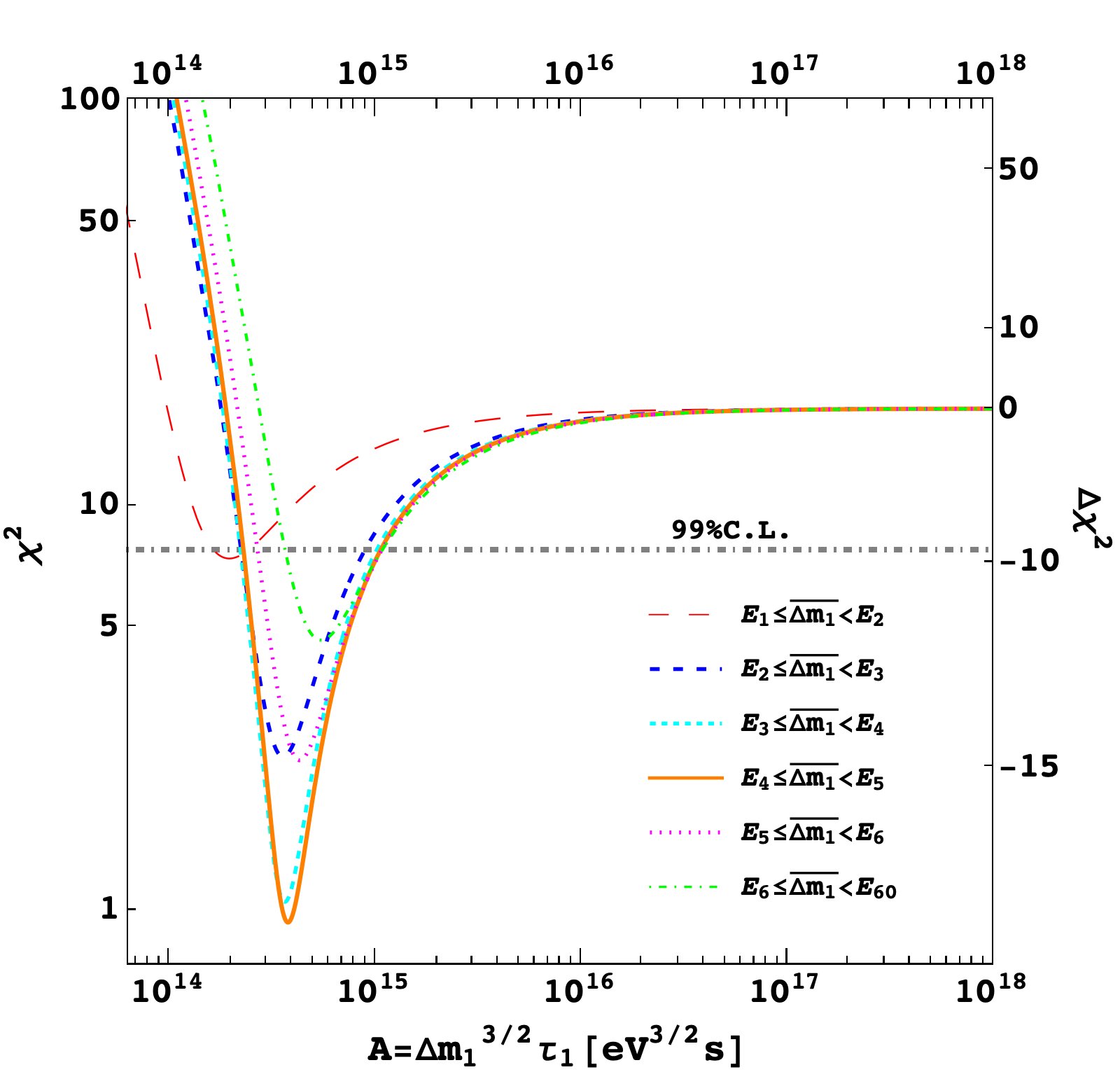,height=7.0cm,width=7.0cm}
\end{center}
    \caption{Left panel: $\chi^2$ versus $\Delta m_1$ for best fit value of $A$ in each interval as 
    in Table 2. Right panel: $\chi^2$ (4 d.o.f.) as a function of $A$ for the best fit values of $\D m_1$ in each interval $E_i \leq \D m_1 < E_{i+1}$ as in Table~\ref{default}.}
    \label{fig:chisq}
\end{figure}

In the case of ARCADE 2, the measurements are made at $z =0$, so that, specialising Eq.~(\ref{Inth}) at $z=0$,
the predicted specific intensity of the non-thermal component produced by the relic neutrino decays is given by 
\be
I_{\g_{\rm nth}}(E,0) ={n^{\infty}_{\nu_1}(0)\over 4\,\pi}\,{e^{-{t(a_{\rm D})\over \tau_1}} \over H(a_{\rm D}) \, \tau_1}   \,  .
\ee
Since we are in the regime $E \ll T_{\gamma}$, we can simply use Eq.~(\ref{Tgammabis}) to calculate the 
predicted effective temperature of the excess radio background from relic neutrino decays. 
Moreover, we will consider the case $\tau_1 \gg t_0$,\footnote{We will comment in the conclusions on the existence of the solution for $\tau_1 \ll t_0$ reported in \cite{Chianese:2018luo}.}
so that the exponential can be safely approximated by unity and in the end we obtain:
\be\label{Tgammanth}
T_{\gamma_{\rm nth}}(E,0) \simeq  {6\,\zeta(3)\over 11 \, \sqrt{\Omega_{{\rm M}0}}}\,
{T_{0}^3 \over E^{1 / 2}\, \Delta m_1^{3 / 2} } \, {t_0 \over \tau_1} 
\,\left(1 +{a_{\rm D}^3 \over a_{\rm eq}^3} \right)^{-{1 \over 2}} \,  .
\ee
In Table~\ref{table1} we show the  effective temperature of the excess radio background measured by ARCADE 2 at 9 different frequencies.
As one can see from the table, only the first four measurements indicate some excess with non-negligible statistical significance. The
fifth and sixth points seem to be compatible with zero within the errors. 
The other three points, seventh to ninth in the table, have very large errors and/or give a negative result.
 As reported by the ARCADE 2 collaboration, this is because they are dominated by noise and for this  reason we just simply disregard them. 
 Moreover, we have placed an upper limit $\D m_1 < 2.5 \times 10^{-4}\,{\rm eV}$,  corresponding to frequencies $\nu < 60 \,{\rm GHz}$, since at higher 
frequencies there are very stringent constraints from the FIRAS measurements \cite{2002ApJFIRAS}. 
\begin{table}[htp]
\caption{ARCADE 2 measurements of the excess radio background effective temperature \cite{Fixsen:2009xn}.}
\begin{center}
\begin{tabular}{c|c|c|c|c|c} \label{table1}
$i$ &$\nu_i$  (GHz) & $E_i$ ($10^{-5}\,{\rm eV}$) & $T^i_{\gamma 0}$  (K)  & $\overline{T}^i_{\rm ERB}$  (mK)   
& $\delta T^i_{\rm ERB}$  (mK)  \\
\hline
 1 & 3.20 & 1.36 & 2.792 & 63 & 10   \\
 2 & 3.41  & 1.41 & 2.771  & 42 & 9  \\
 3 & 7.97 & 3.30 & 2.765 & 36 &  14    \\
 4 & 8.33 & 3.44 & 2.741 & 12 & 16  \\
 5 & 9.72 & 4.02 & 2.732 & 3 & 6  \\
 6 & 10.49 & 4.34 & 2.732 & 3 & 6 \\
 \hline\hline
 7 & 29.5 & 12.2 & 2.529 & $-200$ & 155 \\
 8 & 31 & 12.82 & 2.573 & $-156$ &  76  \\
 9 & 90 & 37.2 & 2.706 & $-23$ & 19  \\
 \hline
\end{tabular}
\end{center}
\label{default}
\end{table}%
Notice that an explanation of the excess radio background detected by ARCADE 2 in term of relic neutrino decays
(or even in terms of any other relic particle decays) necessarily implies $E \leq \D m_1$ and, therefore,
the {\em existence of an endpoint} $E_{\rm max} =\D m_1$. This is a very clear prediction of the model that will be tested
by next experiments aiming at detecting deviations from CMB such as PIXIE, 
with low threshold $\nu_{\rm PIXIE} = 30\, {\rm GHz}$~\cite{Kogut:2011xw} 
and new radio interferometers such as TMS, with low threshold $\nu_{\rm TMS} = 10\,{\rm GHz}$~\cite{2020SPIE11453E..0TR,Alonso-Arias:2021quq}.
This implies that for a fixed value of $\D m_1$,  the model predicts no excess radiation at all energies $E > \D m_1$, so that the standard case is recovered.  For this reason, in each interval $ E_{i} < \D m_1 \leq E_{i+1}$ one has that the $\chi^2(\D m_1, \tau_1)$ varies continuously  with $\D m_1$
but at each borderline energy value $E_i$ one has a discontinuity since the predicted temperature jumps from  some finite
value, for $\D m_1 \leq E_i$, down to zero, for $\D m_1 > E_i$.   Therefore, we determined the $\chi^2_{\rm min}$, and the best fit values, 
separately for each interval. Instead of  $\tau_1$, we have minimised $\chi^2$ as a function of the quantity $A \equiv \D m_1^{3/2}\,\tau_1$
as a first parameter, together with $\D m_1$ as a second parameter. 
This is because, in the matter-dominated regime, $A$ is the only parameter
that determines the effective temperature as one can see from Eq.~(\ref{Tgammanth}).
In the left panel of Fig.~\ref{fig:chisq} we show the $\chi^2$ per degrees of freedom (d.o.f.) as a function of $\D m^2_1$ for a 
fixed value of $A = \bar{A}_i$ in each interval $E_i \leq \D m_1 < E_{i+1}$, 
where $\bar{A}_i$ is the best fit value in that interval.   
One can clearly notice the jumps at each value $E_i$. For $\D m_1 < E_1$, one recovers the value 
$\chi^2_{\rm null}= 17.2$ (4 d.o.f.) corresponding to the standard case, since there is no observable quantity 
that can distinguish between the two cases. 
\begin{table}[t!]
\caption{Results of the fit of ARCADE 2 data. Best fit values, $\chi^2$ and $\D \chi^2$ are shown for each interval of $\D m_1$, corresponding to a frequency interval between two data points.}
\begin{center}
\begin{tabular}{|c|c|c|c|c|c|} 
\hline 
 {\rm Interval} &  $\overline{A} \, ({\rm eV}^{3/2}\,{\rm s})$   & $\overline{\D m}_1$ (eV) & $\overline{\tau}_1$ (${\rm s}$) & $\chi^2_{{\rm min}}$ & $\Delta \chi^2_{{\rm min}}$ \\ [2pt]
\hline 
$ E_1 \leq \D m_1 < E_2$ &  $1.9 \times 10^{14}$ & $1.4 \times 10^{-5}$ & $3.6 \times 10^{21}$ & 7.36 & $-9.87$  \\
 $E_2 \leq \D m_1 < E_3$  &  $2.3 \times 10^{14}$ & $2.7 \times 10^{-5}$ & $1.6 \times 10^{21}$ & 2.28 & $-14.95$ \\
$ E_3 \leq \D m_1 < E_4$  & $3.6 \times 10^{14}$ & $3.4 \times 10^{-5}$ & $1.8 \times 10^{21}$ & 1.06 & $-16.17$   \\
$ E_4 \leq \D m_1 < E_5$  &  $3.8 \times 10^{14}$ & $4.0 \times 10^{-5}$ & $1.46 \times 10^{21}$ & 0.96 & $-16.27$  \\
$E_5 \leq \D m_1 < E_6$  &  $4.2 \times 10^{14}$ & $4.3 \times 10^{-5}$ & $1.49 \times 10^{21}$& 2.19& $-15.04$ \\
$E_6 \leq \D m_1 < E_{60}$  & $4.7 \times 10^{14}$ & $2.0 \times 10^{-4}$ & $1.66 \times 10^{20}$ & 3.23 & $-14.00$ \\
\hline
\end{tabular}
\end{center}
\label{default} 
\end{table}%
The results of the fit are shown in Table~\ref{default} for each interval: we show best fit values for $A$, $\D m_1$ and $\tau_1$
and corresponding $\chi^2_{{\rm min}}$.

\begin{figure}
\begin{center}
  \includegraphics[scale=0.75]{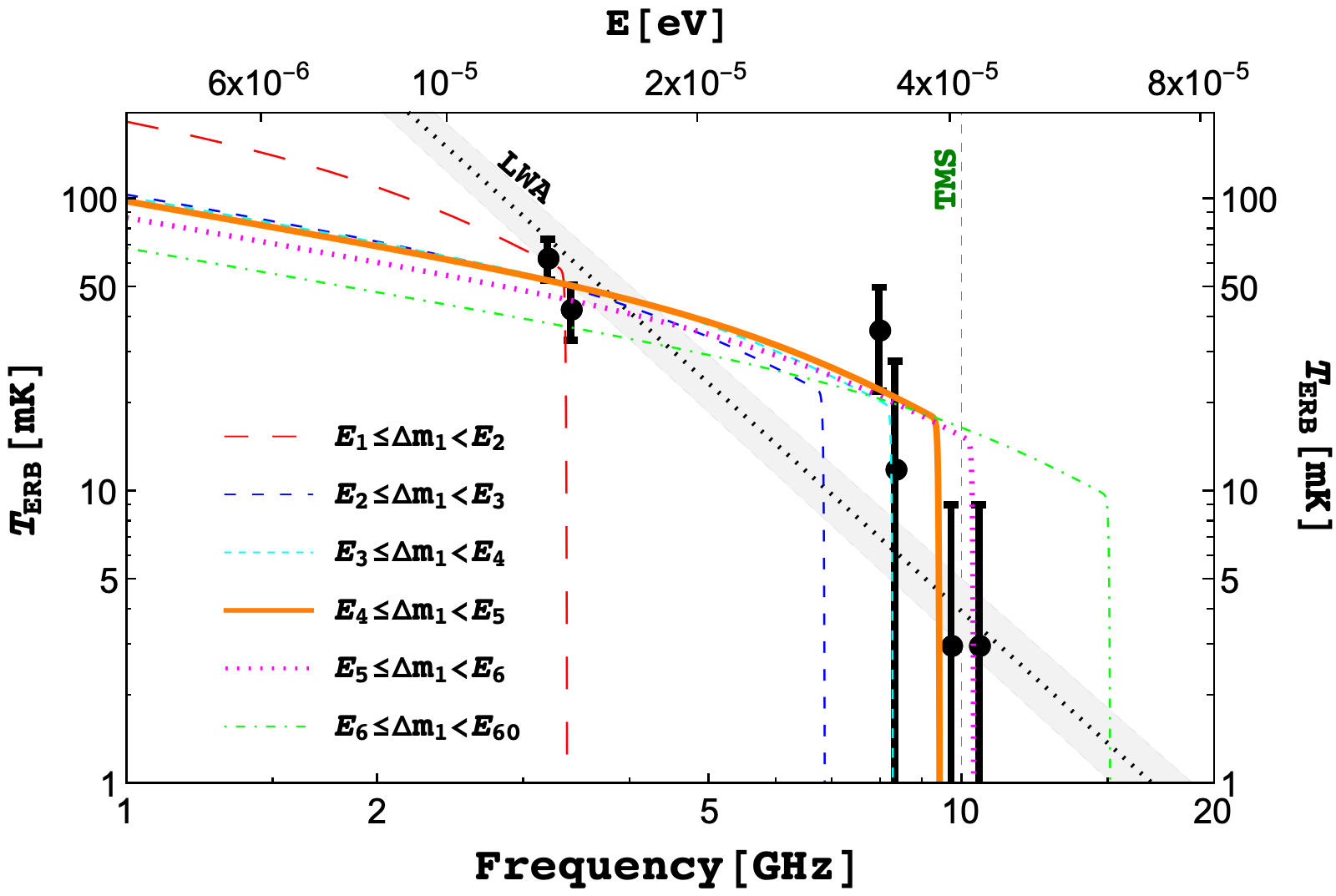}
\end{center}
    \caption{Best fit curves for $T_{\rm ERB}$ obtained with Eq.~(\ref{Tgammanth}). The thick solid orange curve corresponds to a solution very close to the
    best global fit ($\D m_1 = 4.0 \times 10^{-5}\,{\rm eV}$  and $ \tau_1= 1.46 \times 10^{21}\,{\rm s}$).
    The ARCADE 2 data points are taken from Ref.~\cite{Fixsen:2009xn}, while the power-law fit 
    $\beta = -2.58 \pm 0.05$ (dotted line with grey shade) is from \cite{Dowell:2018mdb}.}
    \label{fig:bestfit}
\end{figure}

\begin{figure}
    \centering
    \includegraphics[scale=0.6]{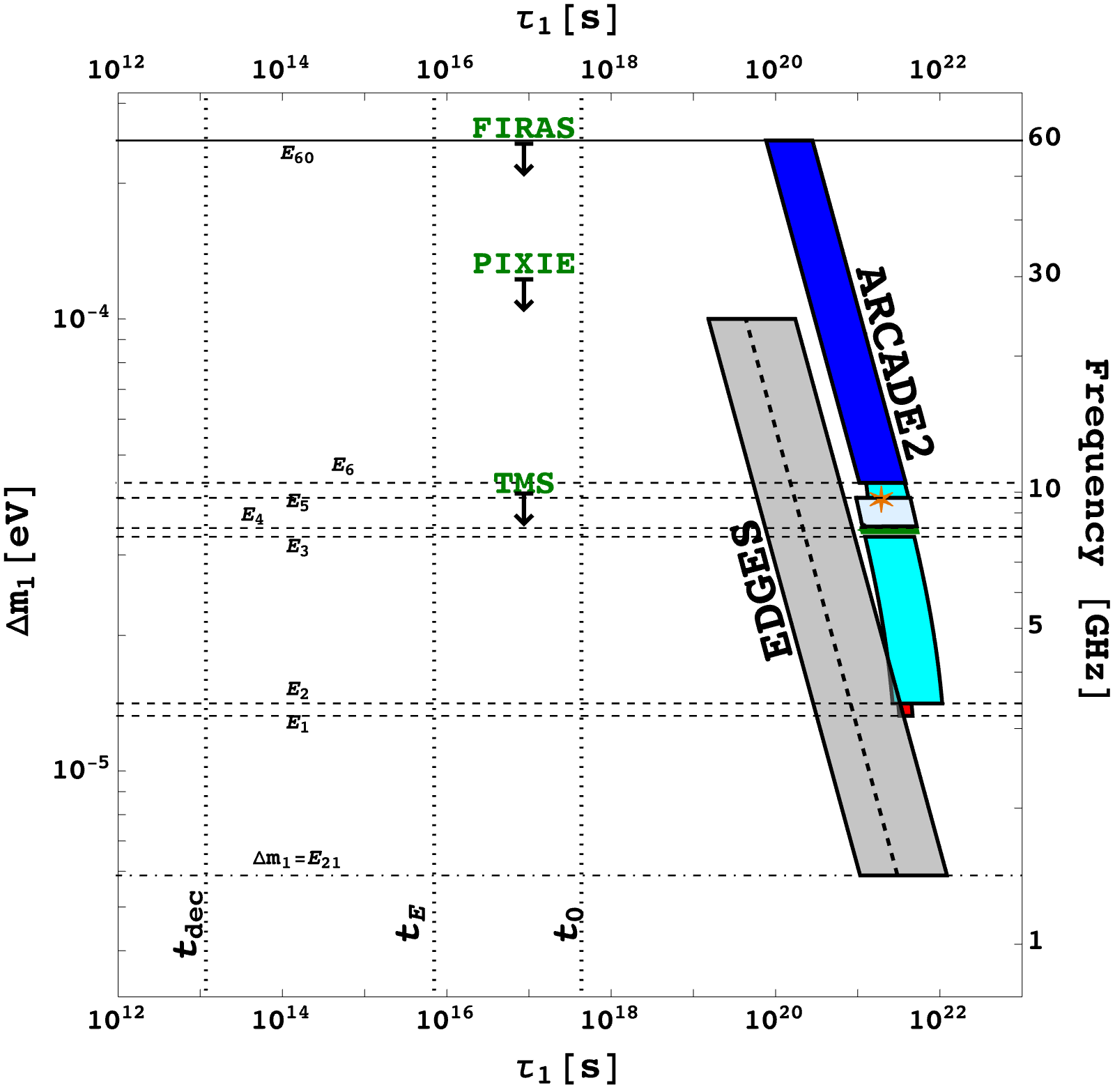}
    \caption{Allowed region in the plane of $\D m_1$ versus $\tau_1$ ($99 \%$ C.L.). The orange star indicates 
    the best fit to ARCADE 2 excess found. Colour code: light grey for $\chi^2 < 1$, green for $1 < \chi^2 < 2$, 
cyan for $ 2 < \chi^2 <3$, blue for $ 3<\chi^2 < 7$ and red  for $7 <\chi^2 < 9$. We also show the best-fit and 99\% C.L. region for the EDGES anomaly (dashed line and grey shaded region). The vertical lines show (from left to right) the epochs of CMB decoupling, cosmic dawn and the age of the Universe. The horizontal lines correspond to different photon energies (or frequencies). The region above 60 GHz is disfavored by FIRAS, and this limit can be improved down to 30 GHz with PIXIE, or to 10 GHz by TMS, as shown by the downward arrows.}
    \label{fig:allowed}
\end{figure}

In the right panel of Fig.~\ref{fig:chisq} we show the $\chi^2$ per d.o.f. as a function of $A$ for the best fit value of $\D m_1$ in each interval as indicated.
For sufficiently large values of $A$ (i.e., large values of $\tau$), one again always recovers the standard case
with $\chi^2_{\rm null} = 17.2$ (4 d.o.f.). In Table~\ref{default} we also
show the different values $\D \chi^2_{{\rm min},i} = \chi^2_{{\rm min},i} - \chi^2_{\rm null}$. 
It is intriguing that the data points seem to indicate the presence of an end point falling  in the interval 
$E_4 < E \sim 4 \times 10^{-5}\,{\rm eV} \leq E_5$ ($\nu_4 < \nu \simeq 9.7\,{\rm GHz} < \nu_5$) as also confirmed by
the results of the fit: the best fit is found indeed for $E_4 < \D m_1 \lesssim E_5 = 4.02 \times 10^{-5}\,{\rm eV}$ with
$\chi^2_{{\rm min},4} = 0.96$. 
In the right panel of Fig.~\ref{fig:bestfit}  we show the best fit curves for $T_{\gamma_{\rm nth}}(E,0)$ for each interval together
with the six ARCADE 2 data points. The thick solid (orange) curve corresponds to the global best fit and it should be clear how
well it fits all data points.
In Fig.~\ref{fig:allowed} we show the resulting $99\%\, {\rm C.L.}$ allowed region in the plane of $\D m_1$ versus $\tau_1$. The different 
blue shades are indicative of the goodness of the fit.
The orange star denotes the global best fit.

If we focus on the interval  $E_4 \leq \D m_1 < E_5 $, where the best fit is found,  
we find $\overline{A} \simeq 3.8 \times 10^{14}\,{\rm eV}^{3/2}\,{\rm s}$ with 
$\chi^2_{\rm min}/{\rm d.o.f.} \simeq 0.96$ (4 d.of.). For $\chi^2 \lesssim \chi^2_{\rm min} + (2.6)^2$, 
(corresponding approximately to the $99\% \, {\rm C.L.}$ range), we find 
\be\label{ARCADE}
A = 3.8^{+7.2}_{-1.5}  \times 10^{14}\,{\rm eV}^{3/2}\,{\rm s} \,  .
\ee
If one interprets the ARCADE 2 excess as due to some alternative explanation than relic neutrino decays, 
then  (\ref{ARCADE}) places the $99\%$ C.L. lower bound
\be 
(\Delta m_1)^{3/2} \, \tau_1 > 2.3 (1.3) \times 10^{14}\,{\rm eV}^{3/2}\,{\rm s} \,  ,
\ee
that, as one can see from Fig.~\ref{fig:allowed},  approximately holds in the range 
$2.5 \times 10^{-5}\,{\rm eV} < \Delta m_1 < E_{60} = 2.5 \times 10^{-4}\,{\rm eV}$
($E_1 = 1.36 \times 10^{-5}\,{\rm eV} \leq \Delta m_1 \leq 2.5 \times 10^{-5}\,{\rm eV}$).
The ARCADE 2 data can be combined with additional data also providing evidence for 
an excess radio background at lower frequencies ($40\, {\rm MHz }\lesssim \nu \lesssim 1.42\,{\rm GHz}$
and in this case a power law yields a very good combined fit \cite{Fixsen:2009xn}.  
More recently, the Long Wavelength Array (LWA) measured the 
diffuse radio background between 40 and 80 MHz, also finding an excess. In
combination with the data from ARCADE 2 and other experiments, the following power law spectrum 
has been found \cite{Dowell:2018mdb}
\be\label{pl}
T_{\rm ERB} = (30.4 \pm 2.6)\,\left({\nu \over 310\,{\rm MHz}} \right)^{-2.58\pm 0.05} \,{\rm K}  \,  .
\ee
This is shown in Fig.~\ref{fig:bestfit} with a grey band. It can be seen however that especially the third point from ARCADE 2
seems not to be well fit by this power law. One indeed finds $\chi^2 \simeq 2.5$ (4 d.o.f.), showing that the power law in Eq.~(\ref{pl}) 
gives a worse fit to the ARCADE 2 data than the best fit we found, but it is still acceptable. However, the power law in Eq.~(\ref{pl})
still requires some physical source that produces it. It is then legitimate to wonder whether it is possible to combine 
the success of relic neutrino decays in reproducing the ARCADE 2 data with the success of the power law (\ref{pl}) in reproducing data below 1 GHz.
As recently pointed out in \cite{Acharya:2023ygd}, the account of soft photon heating from free-free processes might provide a plausible 
solution to find such a combined description of ARCADE 2 with sub-GHz data.  In this case the power law would hold up to 
$\nu \sim 1\,{\rm GHz}$ but should turn into Eq.~(\ref{Tgammanth}) at higher frequencies. Testing such a hypothesis of course would require
a new experiment able to collect data points in the 1 GHz--20 GHz range. The TMS experiment will only take data above 10 GHz. This will be certainly
still useful to strengthen the hint for the existence of an endpoint and possibly rule out 
the validity of the power law in Eq.~(\ref{pl}) up to such high frequencies.  However, it will not provide a thorough test of the best fit solution 
we found to the excess radio background. 
This could come from a new balloon experiment, an improved version of ARCADE 2. Of course an accurate and precise agreement with
the prediction  from Eq.~(\ref{Tgammanth}) could provide a strong evidence (if not a `smoking gun'). 

\section{Predicting the 21 cm cosmological global absorption signal}\label{sec:4}

An additional `smoking gun test' of our relic neutrino decay hypothesis, which predicts non-thermal photons 
also at higher redshifts, can be provided by the observation of the 21 cm cosmological global absorption signal. 
The 21 cm hyperfine transition of the hydrogen atom can be used to test the physics of the cosmic dawn, 
when first astrophysical sources start to form~\cite{Pritchard:2011xb}.  It is a transition between the spin-singlet and
triplet levels of the 1s ground state of the hydrogen atom. The energy gap at rest, giving the energy of the absorbed or emitted photons in the transition, is $E_{21} = 5.87\,\m{\rm eV}$ corresponding to a 21 cm line rest frequency 
 $\nu_{21}^{\rm rest} = 1420\,{\rm MHz}$. The ratio of the population of the excited (spin-triplet) state, $n_1$,
to that one of the ground  (spin-singlet) state, $n_0$, can be parameterised in terms of a {\em spin temperature} 
$T_{\rm S}(z)$ depending on the redshift:
\be
{n_1 \over n_0}(z) \equiv {g_1 \over g_0} \, e^{-{E_{21} \over T_{\rm S}(z)}} \,  ,
\ee 
where $g_1/g_0 = 3$ is the excited state-to-ground state degeneracy ratio.
The possibility to detect an observable signal from the 21 cm cosmological transitions of the hydrogen
atoms in the primordial gas after recombination relies on a non-vanishing {\em brightness constrast} 
between the cosmic radiation and the radiation emitted in the 21 cm transitions. This can be parameterised
in terms of the 21cm  brightness temperature $T_{21}(z)$ \cite{Zaldarriaga:2003du}
\be\label{T21}
T_{21}(z) \simeq  23\,{\rm mK} \, (1+\d_{\rm B})\, x_{H_I}(z) \,\left({\Omega_{{\rm B}0} h^2 \over 0.02}\right)\,
\left[\left({0.15 \over \Omega_{{\rm M}0}h^2}\right)\,
\left({1+z \over 10} \right)  \right]^{1/2}  \,\left[
1 - {T_\g(z) \over T_{\rm S}(z)} \right]  \,  ,
\ee
where $\d_{\rm B} = (\rho_{\rm B}-\bar{\rho}_{\rm B})/\bar{\rho}_{\rm B}$ is the fractional
baryon overdensity, $x_{H_I}(z)$ is the neutral hydrogen fraction, $\Omega_{{\rm B}0}$
and $\Omega_{{\rm M}0}$ are the baryon and matter energy density parameters, $h$ is the Hubble
constant in units of $100\,{\rm km}\,{\rm s}^{-1}\,{\rm Mpc}^{-1}$.
If the spin temperature is equal to the photon temperature ($T_{\rm S} = T_{\gamma}$), 
then photons are absorbed and reemitted with the same intensity and there is no visible signal. 
Also, if all atoms are ionised so that $x_{H_I} = 0$, there cannot be any signal. Another important point
to consider is that the spin temperature is related to $T_{\rm gas}$, the kinetic temperature of the gas,  by
\be
\left(1- {T_{\gamma} \over T_{\rm S}}\right) \simeq {x_c + x_\a \over 1 + x_c + x_\a }\, \left(1- {T_{\gamma} \over T_{\rm gas}}\right)   \,   , 
\ee
where $x_\a$ and $x_c$ are coefficients describing the coupling between 
the hyperfine levels and the gas.  
Specifically, the $x_c$ term is for collisional excitation of the hyperfine transition, and the $x_\alpha$ term is 
for the Wouthuysen-Field effect, mediated by the absorption and subsequent re-emission of a Lyman-$\alpha$ photon.
In the limit of strong coupling, for $x_\a + x_c \gg 1$,
one has $T_{\rm S} = T_{\rm gas}$, while in the limit of no coupling, for $x_\a = x_c =0$, one has 
$T_{\rm S} = T_{\gamma}$ and in this case also there is no signal.  

From these considerations,
one can then draw the following evolution for the 21cm global signal \cite{Furlanetto:2006jb,Pritchard:2011xb}:
\begin{itemize}
\item[(i)] The hydrogen gas decouples completely from the cosmic radiation, so that one can have $T_{\rm S} \neq 
T_{\gamma}$, at a redshift $z_{\rm dec}  \simeq 150$. For lower redshifts, the gas cools down more
rapidly than cosmic radiation, with $T_{\rm gas}(z) = T(z_{\rm dec}) \, (1 + z_{\rm dec})^2$. Moreover
in the redshift range $z_{\rm dec} \gtrsim z \gtrsim 30$, the gas collision rate is high enough that the 21 cm transitions
are coupled to the gas temperature and approximately $T_{\rm S} \simeq T_{\rm gas}$. During this stage 
an absorption signal, for $T_{21} < 0$, is expected.  
\item[(ii)] At redshifts $z \simeq 30$, the gas becomes so rarified that 
collisions are unable to couple the spin temperature to the gas temperature and one has 
$x_a + x_c \ll 1$ and $T_{\rm S} \simeq T_{\gamma}$. Therefore, at this stage $T_{21} \simeq 0$ and there is no 21 cm signal again.
\item[(iii)] At redshifts $z_\star \simeq 30$,  first luminous sources also start to form and this, via Wouthuysen-Field effect, triggers a gradual re-coupling of the 21 cm transitions with the gas so that the spin temperature gradually tends again to the gas temperature. In this stage
one can again expect an absorption signal since $T_{21}< 0$.  
\item[(iv)] Finally, for redshifts around $z \simeq 10$, the gas gets reheated by the
astrophysical radiation and the gas temperature increases. In this stage it might even be possible to 
have $T_{\rm S} \simeq T_{\rm gas} > T_\gamma$, so that one has an emission signal from regions where the gas is not fully ionised.
All gas gets eventually ionised, so that the signal switches off again. 
\end{itemize}
The EDGES collaboration has found an absorption profile in the range $z =15$--$20$, with the minimum at
a frequency $\nu_{21}(z_E) \simeq 78\,{\rm MHz}$ corresponding to $z_E  \simeq 17$ \cite{Bowman:2018yin}. 
The absorption profile has a U shape with quite a flat minimum.  We do not attempt at fitting the precise shape of the profile 
since this may be sensitive to systematic errors and the assumed form of the fitted model. 
However, we will take into account that the exact frequency of the minimum and, consequently, its redshift
has an uncertainty and one can actually write $z_E = \bar{z}_E \pm 1$. Let us first focus   
 on the central value $\bar{z}_E = 17$. At this value, the $\Lambda$CDM model predicts a relic photon temperature
\be
T_{\gamma}(\bar{z}_E) = T(\bar{z}_E) = T_0 \, (1+\bar{z}_E) \simeq 49.6 \, {\rm K} \,  ,
\ee
and a gas temperature $T_{\rm gas}(\bar{z}_E) \simeq 7.2 \, {\rm K}$.\footnote{
Assuming an instantaneous decoupling, one can approximately estimate the gas temperature as
\be
T_{\rm gas}(\bar{z}_E) \simeq T(z_{\rm dec}) \, {(1 + \bar{z}_E)^2 \over (1 + z_{\rm dec})^2} \simeq 6\,{\rm K} \,  .
\ee
The more accurate result in the text comes from a  solution of kinetic equations  \cite{Fialkov:2019vnb}.
}
Defining  $\xi(z) \equiv T_{\gamma}(z)/T_{\rm gas}(z)$, one has then $\xi(\bar{z}_E) \simeq 6.89$, corresponding, to $T_{21}(\bar{z}_E) \simeq -206\,{\rm mK}$ from Eq.~(\ref{T21}).\footnote{We  
are using $\Omega_{{\rm M}0}h^2=0.1424$ and $\Omega_{{\rm B}0}h^2 = 0.02242$ \cite{Planck:2018vyg}.}

On the other hand, the EDGES collaboration finds a much stronger absorption signal with
$T_{21}^{\rm EDGES}(z_E) = -500^{+200}_{-500}\,{\rm mK}$ ($99\%\,{\rm C.L.}$). 
From Eq.~(\ref{T21}), this translates into 
\be
\xi^{\rm EDGES}(\bar{z}_E)=15.1_{-5.6}^{+14.1} \,  ,
\ee 
and from this one finds first   
\be
T^{\rm EDGES}_{\gamma}(\bar{z}_E) = \xi^{\rm EDGES}(\bar{z}_E)\, T_{\rm gas}(\bar{z}_E) =109^{+101}_{-41} \,  {\rm K} \,  ,
\ee
and then finally, subtracting the temperature of thermal photons,  one obtains for 
the effective temperature of the non-thermal contribution ($99\%\,{\rm C.L.}$) 
\be\label{TEDGES}
T^{\rm EDGES}_{\gamma_{\rm nth}}(\bar{z}_E) = (60^{+101}_{-41})\,{\rm K} \,  .
\ee
Finally, we have also calculated the additional error coming from the uncertainty $\Delta z_E = \pm 1$ and found that this  is of order of $3\,{\rm K}$ so that, once added in quadrature, it gives a negligible increase of $0.1\,{\rm K}$ to the total error.

Eq.(\ref{TEDGES}) is the experimental quantity we need to reproduce. We have now to use Eq.~(\ref{Tgammabis}) to calculate the effective
temperature from the specific intensity and Eq.~(\ref{Inth}) for the specific intensity of photons produced from relic neutrino decays. 
This time, in the case of EDGES, the redshift is $z_E$ and the energy is $E_{21}$, so that we obtain
\be\label{TgammanthEDGES}
T_{\gamma_{\rm nth}}(E_{21},z_E) \simeq  {6\,\zeta(3)\over 11 \, \sqrt{\Omega_{{\rm M}0}}}\,
{T_{0}^3\,(1 +z_E)^{3/2}\over E_{21}^{1 / 2}\, \Delta m_1^{3 / 2} } \, {t_0 \over \tau_1}  \,  ,
\ee
where we have again neglected the exponential, considering the case $\tau_1 \gg t(z_{E})$, and we
used the approximation $H(a_{\rm D}) \simeq H_0\,\sqrt{\Omega_{M0}}\,(1+z_E)^{3/2}\, (\D m_1/E_{21})^{{3\over 2}}$, 
holding in the matter-dominated regime. Imposing that this 
predicted effective temperature reproduces  the experimental measurement in Eq.~(\ref{TEDGES}), one finds:
\be
A^{\rm EDGES}(\bar{z}_E) = {6\,\zeta(3)\over 11 \, \sqrt{\Omega_{{\rm M}0}}}\,
{T_{0}^3\,(1 +\bar{z}_E)^{3/2}\over T^{\rm EDGES}_{\gamma_{\rm nth}}(\bar{z}_E) \, E_{21}^{1 / 2} } \, t_0  = 
(4.0 ^{+8.8}_{-2.5}) \times 10^{13}\, \,{\rm eV}^{3/2}\,{\rm s} \;\;\; (99\%\,{\rm C.L.}) .
\ee
This result holds in the range $E_{21} = 5.87\,\m{\rm eV} \leq \D m_1 \lesssim 10^{-4} \, {\rm eV}$. The lower bound is clear since
only photons produced with energies above $E_{21}$ at redshifts $z_{\rm D} \geq z_E$ 
can excite the hydrogen atom ground states. The upper bound is due to the fact that we are treating 
neutrinos as non-relativistic at decay. Therefore, one has to impose $T_{\nu}(z_{\rm D}) \lesssim m_1/3 \lesssim 0.05\,{\rm eV}$,
where in the last inequality we have used the upper bound $m_1 \lesssim 0.15\,{\rm eV}$ coming 
from {\em Planck} data \cite{Planck:2018vyg}.\footnote{We are using a conservative upper bound on the sum of the neutrino masses ($95\% \, {\rm C.L.}$) \cite{Planck:2018vyg}
\be
\sum_i m_i < 0.44 \,{\rm eV} \,  ,
\ee
derived just from {\em Planck} data (no BAO), combining temperature anisotropies (TT), low E polarisation data and lensing. 
The use of such conservative bound is justified by the existing cosmological tensions, that support cosmological models
beyond the minimal $\Lambda$CDM model resulting in a relaxation of the upper bound on neutrino masses.}
This implies $1+z_{\rm D} \lesssim 300$ and consequently $\D m_1 \lesssim E_{21} (1+z_{\rm D})/(1+z_E) \sim 10^{-4}\,{\rm eV}$.

The resulting ($99\%\,{\rm C.L.}$) allowed region  in the plane of 
$\D m_1$ versus $\tau_1$ is shown in  Fig.~\ref{fig:allowed} in light grey.
This has to be compared with the allowed region obtained fitting ARCADE 2 data in Eq.~(\ref{ARCADE}).
  It is clear that the two regions only marginally overlap and so there is some tension between the two results. This tension might be 
reconciled if the current EDGES measurement is overestimating the 21 cm global absorption signal. 

Reversely, it is easy to derive a prediction for the contrast brightness temperature $T_{21}$, that future 21 cm cosmology experiments should measure, starting from our ARCADE 2 data fit.  
This can be simply done plugging first Eq.~(\ref{ARCADE}), giving the ($99\%\,{\rm C.L.}$) allowed range for $A$ from ARCADE 2 data at the best fit value for $\D m_1$, into Eq.~(\ref{TgammanthEDGES}), finding $T_{\gamma_{\rm nth}}(E_{21},z_E)$, and then from Eq.~(\ref{T21}) one obtains:
\be\label{T21predicted}
T_{21}(\bar{z}_E) = -238^{+21}_{-20}\,{\rm mK} \;\;\; (99\%\,{\rm C.L.}) . 
\ee  
When this is compared with the EDGES measurement, unsurprisingly, one again finds the same kind of 
tension found in terms of $A$, now expressed in terms of the observable $T_{21}$. 
The result (\ref{T21predicted}) is one of our main results, since it is another clear prediction of the model.
This, together with a test of the predictions of the excess radio background effective temperature, represents a clear signature of the model. 

\section{On dark matter decays/de-excitations as alternative explanation}\label{sec:5}

The solution we have found can also be mimicked by replacing relic neutrino decays with dark matter decays or de-excitations. 
This kind of models was proposed as explanation the EDGES anomaly \cite{Fraser:2018acy}.\footnote{In this case, 
one should use Eq.~(\ref{Inth}) with $n^{\infty}_{\nu_1}(z_E)$ replaced by $n^{\infty}_{\rm DM}(z_E)$,
the number density of excited dark matter at the redshift $z_E$. However, in \cite{Fraser:2018acy} 
the number density of excited dark matter at the present time,  $n^{\infty}_{{\rm DM},0}$, was used, as in the case of 
ARCADE 2 excess radio background, that is a factor $(1+z_E)^3$ smaller.}
For example, one can have in mind a picture where dark matter is made of some dark atoms and photons are
produced from de-excitations of the first excited state into the ground state. 
Indicating with $m_{\rm DM}$
the mass of dark matter and with $f_{\rm D}$ the fraction of dark matter abundance in the excited state,
we obtain for the effective temperature of the excess radio background
\be
T^{(\rm DM)}_{\g_{\rm nth}}(E,0) ={\pi^2\,f_{\rm D}\,n^{\infty}_{\rm DM}(0)\over E^2}\,{e^{-{t(a_{\rm D})\over \tau_{\rm DM}}} \over H(a_{\rm D}) \, \tau_{\rm DM}}   \,   .
\ee 
The (average) dark matter number density at the present time is simply given by  
\be
n^{\infty}_{\rm DM}(0) = {\rho_{{\rm c}0}h^{-2} \over m_{\rm DM}} \, \Omega_{{\rm DM}0}h^2 \simeq 1.26 \, {{\rm GeV} \over m_{\rm DM}}\,{\rm m}^{-3} \,  , 
\ee
where we used $\Omega_{{\rm DM}0}h^2 = 0.11933$ \cite{Planck:2018vyg} 
and $\rho_{{\rm c}0}h^{-2} =10.54\,{\rm GeV}\,{\rm m}^{-3}$,
with $\rho_{{\rm c}0}$ the critical energy density at the present time. 
In terms of the neutrino number density $n_{\nu_1}^{\infty}(0)$ (see Eq.~(\ref{nnu})), it can also be rewritten as 
\be
n^{\infty}_{\rm DM}(0) = {11.3 \, {\rm eV} \over m_{\rm DM}} \, n_{\nu_1}^{\infty}(0) \,  .
\ee 
In this way the effective temperature of the excess radio background from dark matter
de-excitations can be expressed in terms of the one we found for relic neutrino decays
(Eq.~(\ref{Tgammanth})):
\be\label{TDMg}
T^{(\rm DM)}_{\g_{\rm nth}}(E,0) =\left({11.3\,{\rm eV}\over m_{\rm DM}\,f^{-1}_{\rm D}}\right)
\,{{\tau}_1 \over \tau_{\rm DM} }
\, T^{(\nu)}_{\g_{\rm nth}}(E,0)\, e^{-{t(a_{\rm D})\over \tau_{\rm DM}}} \,  .
\ee 
If we approximate $t(a_{\rm D}) \simeq t_0$, strictly valid for $E = \D m_1$, then it is 
clear that the same solution we obtained for neutrinos is also obtained for 
dark matter decays/de-excitations replacing $\tau_1 \ra \tau_{\rm DM}$ (and of course
$\D m_1 \ra \D m_{\rm DM}$) for 
\be
m_{\rm DM}\,f^{-1}_{\rm D} \sim 10\,{\rm keV}  {t_0 \over \tau_{\rm DM}} \, e^{-{t_0 \over \tau_{\rm DM}}} \,  .
\ee
As long as $\tau_{\rm DM} \gtrsim t_0$, the exponential does not play any role.
One can see that  for $m_{\rm DM}/f_{\rm D} \sim 10\,{\rm eV}$, one 
obtains the same value of the lifetime we obtained for neutrinos, $\tau_{\rm DM} \sim \bar{\tau}_1 \sim 10^{21}\,{\rm s}$. For increasing values of $m_{\rm DM}/f_{\rm D}$, the lifetime, and therefore the value of $A$, decreases as
$\tau_{\rm DM} \propto (m_{\rm DM}/f_{\rm D})^{-1}$. 
For $m_{\rm DM}/f_{\rm D} \sim 10\,{\rm keV}$, one has $\tau_{\rm DM}\sim t_0$. In this case,
the exponential kicks in and the effective temperature starts to be suppressed exponentially. This occurs because 
the dark matter number density, and correspondingly the specific intensity, starts to become too low.  
Therefore, the ARCADE 2 data cannot be reproduced for values of $m_{\rm DM}/f_{\rm D}$ much larger than $10\,{\rm keV}$ and one obtains the approximate upper bound
\be\label{upan}
m_{\rm DM} \lesssim 10\,{\rm keV} \, f_{\rm D} \,   .
\ee  
Notice though that one can still reproduce the EDGES data since in that case one has to replace 
$t_0 \ra t_E = t(\bar{z}_E) \simeq 250\,{\rm Myr}\sim 10^{16}\,{\rm s}$. In this case the upper bound relaxes  by about  two orders of magnitude. Of course in this case it would be impossible to reconcile EDGES anomaly
and ARCADE 2 excess radio background. We have also derived this upper bound numerically, using Eq.~(\ref{TDMg}) to fit the
ARCADE 2 data, calculating the $\chi^2$, as we did in the case of neutrino decays, for a few values of 
$m_{\rm DM}/f_{\rm D}$.  In this case, however, the exponential has to be taken into account, since it plays a crucial role in placing the upper bound.  Fixing $\D m_{\rm DM} = 4\times 10^{-5}\,{\rm eV}$, the best fit value obtained for $\D m_1$ in the case of
relic neutrino decays, we show in Table \ref{table3} the values of $\chi^2_{\rm min}$ (4 d.o.f.) and $\bar{\tau}_{\rm DM}$ we have obtained for some (increasing) values of $m_{\rm DM}/f_{\rm D}$.

\begin{table}[htp]
\caption{Values of $\chi^2_{\rm min}$ and $\bar{\tau}_{\rm DM}$ for some increasing values of
$m_{\rm DM}/f_{\rm D}$.}
\begin{center}
\begin{tabular}{|c|c|c|} 
\hline 
$m_{\rm DM}/f_{\rm D} $ & $\chi^2_{\rm min}$ & $\bar{\tau}_{\rm DM}$ \\
\hline
11.3 eV & 0.96 &  $1.46 \times 10^{21}$ s \\
100 eV & 0.97 &  $1.67 \times 10^{20}$ s \\
1 keV & 0.98 &  $1.67 \times 10^{19}$ s \\
10\,{\rm keV} & 1.15 & $1.68 \times 10^{18}\,{\rm s}$ \\
100\,{\rm keV} & 6.6 & $1.4 \times 10^{17}\,{\rm s}$ \\
\hline
\end{tabular}
\end{center}
\label{table3}
\end{table}
As one can see,  for $m_{\rm DM}/f_{\rm D}= 11.3 \, {\rm eV}$, one recovers the same best fit obtained for relic neutrino decays.
For $m_{\rm DM}/f_{\rm D} \gtrsim 10\,{\rm keV}$, the fit rapidly deteriorates and one obtains an upper bound
$m_{\rm DM}/f_{\rm D} \lesssim 100\,{\rm keV}$ at about $99\%$ C.L.. For larger values, one recovers the null hypothesis,
since the dark matter number density becomes so low that the predicted values of the effective temperature become much lower than the measured ones for any value of  $\tau_{\rm DM}$.  Notice that, compared to the simple analytical estimate Eq.~(\ref{upan}), the upper bound is somehow relaxed. This is 
because the values $t(a_{{\rm D},i})$, obtained for $a_{{\rm D},i} = E_i/\D m_{\rm DM}$, are one-two orders of magnitude smaller than $t_0$.

 We should also make clear that this upper bound on the dark matter mass is obtained under the assumptions of validity of the solution. In particular, this holds under the assumption that the excess radio background is explained by the primary photons produced from dark matter decays/de-excitations that are simply redshifted from decays
to the present time in case of the excess radio background or to $t_E$ in the case of the EDGES anomaly. 
This does not exclude that other processes, such as soft photon heating, could be active at higher dark matter
mass or mass difference values, reprocessing injected photons and reproducing  the excess radio background, though with a different spectrum, typically a power law one \cite{Acharya:2023ygd}. 
  
Finally, notice that these strong limitations on a solution in terms of dark matter decays/de-excitations are in addition to the strong constraints from the smoothness of the excess radio background that clashes with
the expected anisotropies of the signal that should track the dark matter density distribution.\footnote{In \cite{Holder:2012nm}, a $95 \%$ C.L. upper bound of $1\%$ has been derived on the clustering amplitude at small scales (between 300 kpc  
and 3 Mpc) using ATCA data at 8.7 GHz \cite{Subrahmanyan:2000df}. 
On the other hand, expected clustering amplitude from radio sources tracking matter distribution, as it would be
of course the case of decaying dark matter, is about 5 times higher. This clearly
rules out these models, at least in their minimal version. One has to introduce some
{\em ad hoc} smoothing mechanism to rescue them.} 

\section{Final remarks}\label{sec:6}

Let us conclude with some final remarks on the solution to the excess radio background we have discussed.

\begin{itemize}
\item We have referred to the case of lightest neutrino decays into sterile neutrinos. However, as we said in the introduction,
the discussion is equally applicable to the case when one of the two heavier light neutrinos, either $\nu_2$ or $\nu_3$, 
decays into sterile neutrinos; simply one has to replace everywhere $\D m_1 \rightarrow \D m_{2,3}$ 
and $ \tau_1 \rightarrow \tau_{2,3}$. The difference
is that $m_2$ and $m_3$ are lower-bounded by the solar neutrino and atmospheric neutrino mass scales, respectively, i.e.,
$m_2 \gtrsim 8\,{\rm meV}$ and $m_3 \gtrsim 50\,{\rm meV}$ (in this case $m_1$ can be arbitrarily small). 
This lower bound guarantees they are automatically non-relativistic 
in all the relevant range of redshifts ($0$--$300$).  On the other hand, in the case of lightest active neutrinos, we have to impose 
$m_1 \gtrsim 10\,{\rm meV}$, an interesting condition that might be tested during next years by absolute neutrino mass scale experiments (including cosmology). 
It is also intriguing to notice that since the clustering of neutrinos is proportional to their mass, then in principle
excess radio background anisotropies, either a detection or also an upper bound, 
would allow to distinguish among the three cases and even give information on the absolute neutrino mass scale. 
\item The solution we found predicts an effective temperature of the excess radio background that depends
on photon energy as $\propto E^{-0.5}$
for $a_{\rm D}= E/\D m_i \lesssim a_{\rm eq}\simeq 0.75$ and as $\propto E^{-1.5}$ for $a_{\rm D} \gtrsim 0.75$, with an abrupt endpoint at $E = \D m_i$. These are all features that makes it distinguishable from other solutions and in particular from the power law Eq.~(\ref{pl}).
To this extent measurements in the frequency range $\nu = 1\,{\rm }$--$20\,{\rm GHz}$ would be needed and so an improved ARCADE 2-like 
experiment would be certainly  a priority. The TMS microwave telescope, will only partly test the solution at $\nu \geq 10\,{\rm GHz}$. The best
fit solution we found predicts that no excess will be detected at TMS. On the other hand, one expects that the power law, well describing the excess radio background measurements at low frequencies, should be recovered accounting for processes such as soft photon heating reprocessing lower frequency photons. 
\item As pointed out in \cite{Chianese:2018luo}, there is a second solution for life-times much shorter than the 
age of the universe. In this case the exponential in Eq.~(\ref{Tgammanth}) cannot be neglected.
However, this solution is strongly fine-tuned and it  lives only within a very narrow range of $A$ values. In addition,
for such short life-times, free-free processes should be taken into account. Moreover, this solution predicts
a negligible deviation in the absorption 21 cm cosmological signal  from the standard case, so that the
tension with the EDGES anomaly is even more exacerbated.   
\item The excess radio background is a detection of photons produced by relic neutrino decays at relatively low redshifts (the highest redshift
for the best fit solution is given by $z \simeq E_5/E_1 - 1 \simeq  2$). However, the 21 cm global signal is sensitive also to photons produced at higher redshifts (up to $\sim 300$). In particular, we have seen how solution for the excess radio background predicts $T_{21} \simeq - 240\,{\rm mK}$. This is another very
specific prediction that will be tested during next years.  This prediction is in mild tension with the anomalous EDGES measurement
of the 21 cm absorption global signal. A possible explanation is that this might have been overestimated due to some unidentified foreground. 
\item The same solution we found in the case of relic neutrino decays could also be mimicked  by decays/de-excitations of some fraction $f_{\rm D}$ of (cold/warm) dark matter. However, this solution is only possible
for masses $m_{\rm DM}/f_{\rm D} \lesssim 10\,{\rm keV}$.  This limitation 
is in addition to the difficulty of such solutions to explain the smoothness 
of the excess radio background, since they unavoidably predict anisotropies tracking the dark matter density distribution. 
\item The solution we have presented assumes that the effect of stimulated decay effects, as considered in \cite{Bolliet:2020ofj}, 
is absent.  This can be nicely realised within a model where neutrinos are unstable, 
and will be presented in a forthcoming paper \cite{preparation}. 
\item Our solution relies on the fact that there is a small mass splitting between the active and sterile neutrino components, or in other words, neutrinos are pseudo-Dirac (albeit with non-maximal mixing).\footnote{For concrete ultraviolet-complete model realizations of the pseudo-Dirac hypothesis, see e.g.~\cite{Babu:2022ikf, Carloni:2022cqz}. It is interesting to note that certain string landscape (swampland) constructions predict that neutrinos are necessarily pseudo-Dirac~\cite{Ooguri:2016pdq, Ibanez:2017kvh, Gonzalo:2021zsp}.} The cosmological probes proposed here are complementary to other laboratory~\cite{deGouvea:2009fp, Ansarifard:2022kvy, Chen:2022zts, Franklin:2023diy} and astrophysical~\cite{2002ApJS, 2000ApJS, Keranen:2003xd, Beacom:2003eu, Esmaili:2009fk, Esmaili:2012ac, Martinez-Soler:2021unz, Carloni:2022cqz, Rink:2022nvw} probes of pseudo-Dirac neutrinos. 
\end{itemize}
It is quite exciting that radio background experiments have the opportunity during next years to test the stability of the 
cosmic neutrino background. The relic neutrino decay solution of the excess radio background we have discussed 
clearly relies on new physics and, therefore, it might provide a very important
guidance toward an extension of our current established fundamental theories of nature. 
In particular, it might help resolve the nature of neutrino mass -- one of the fundamental open questions.

\vspace{-1mm}
\subsection*{Acknowledgments}

The work of BD was partly supported by the U.S. Department of Energy under grant No. DE-
SC 0017987.  PDB and RR acknowledge financial support from the STFC Consolidated Grant ST/T000775/1.
IMS is supported by STFC grant ST/T001011/1.
This project has received funding from the European Union’s Horizon Europe research and innovation programme  under   the Marie Sk\l odowska-Curie Staff Exchange  grant agreement
No. 101086085 – ASYMMETRY.
In particular, PDB wishes to thank the MIT Center for Theoretical Physics for the hospitality and Tracy Slatyer for
useful discussions. We also wish to thank Jens Chluba for useful comments. 

\bibliography{ref}

\bibliographystyle{JHEP}

\end{document}